\documentclass[final,aps,pra,twocolumn,amsmath,amssymb,superscriptaddress,floatfix,]{revtex4-2}

\usepackage{graphicx}

\usepackage{dcolumn}
\usepackage{bm}
\usepackage[utf8]{inputenc}
\usepackage[T1]{fontenc}
\usepackage{mathptmx}
\usepackage{etoolbox}
\usepackage{siunitx}
\usepackage[colorlinks=true, urlcolor  = blue,linkcolor=blue, citecolor=blue,final]{hyperref}

\begin{document}

\title{Homogeneous Free-Standing Nanostructures from Bulk Diamond\\ over Millimeter Scales for Quantum Technologies}

\author{Andrea Corazza}
 \affiliation{Department of Physics, University of Basel, CH-4056 Basel, Switzerland}
\author{Silvia Ruffieux}
\affiliation{Department of Physics, University of Basel, CH-4056 Basel, Switzerland}
\author{Yuchun Zhu}
\affiliation{Institute of Physics and Center for Quantum Science and Engineering, Ecole Polytechnique Fédérale de Lausanne (EPFL), 1015 Lausanne, Switzerland}
\author{Claudio A. Jaramillo Concha}
\affiliation{Institute of Physics and Center for Quantum Science and Engineering, Ecole Polytechnique Fédérale de Lausanne (EPFL), 1015 Lausanne, Switzerland}
\author{Yannik Fontana}
\affiliation{Department of Physics, University of Basel, CH-4056 Basel, Switzerland}
\author{Christophe Galland}
\affiliation{Institute of Physics and Center for Quantum Science and Engineering, Ecole Polytechnique Fédérale de Lausanne (EPFL), 1015 Lausanne, Switzerland}
\author{Richard J. Warburton}
\affiliation{Department of Physics, University of Basel, CH-4056 Basel, Switzerland}
\author{Patrick Maletinsky}
\email{patrick.maletinsky@unibas.ch}
\affiliation{Department of Physics, University of Basel, CH-4056 Basel, Switzerland}

\date{June 11, 2025}

\begin{abstract}
Quantum devices based on optically addressable spin qubits in diamond are promising platforms for quantum technologies such as quantum sensing and communication.
Nano- and microstructuring of the diamond crystal is essential to enhance device performance, yet fabrication remains challenging and often involves trade-offs in surface quality, aspect ratio, device size, and uniformity.
We tackle this hurdle with an approach producing millimeter-scale, thin (down to 70\,nm) and highly parallel ($<$\,0.35\,nm/\textmu m) membranes from single-crystal diamond. The membranes remain contamination-free and possess atomically smooth surfaces ($\mathrm{R_q}$\,$<$\,200\,pm) as required by state-of-the-art quantum applications.
We demonstrate the benefits and versatility of our method by fabricating large fields of free-standing and homogeneous photonic nano- and microstructures. 
Leveraging a refined photolithography-based strategy, our method offers enhanced scalability and produces robust structures suitable for direct use, while remaining compatible with heterogeneous integration through pick-and-place transfer techniques.
\end{abstract}

\maketitle

Diamond's unparalleled properties make it an enticing choice for a plethora of applications, including low-loss micro-electro-mechanical systems~\cite{Tao2014, Wu2018, Zhang2025}, high-power electronics and optics~\cite{Zhu2017, Donato2019, Sasama2022, Hasan2024} as well as quantum photonics, phononics, and sensing ~\cite{Schroeder2016, Shandilya2022, Rovny2024}.
Notably, diamond can incorporate a variety of optically-active point defects (color centers) that also exhibit coherent ground-state spins~\cite{Doherty2013, Thiering2020}. The intrinsic spin-photon interface provided by color centers in single-crystal diamond has enabled multiple breakthroughs pertaining to quantum registers and networks~\cite{Bernien2013, Bradley2019, Bhaskar2020, Baier2021, Stas2022}. Additionally, their spin can serve as powerful sensors either as ensembles~\cite{Wolf2015, Barry2020, Scholten2021} or at the single center level~\cite{Devience2015, Thiel2019, Rovny2022}.
At the core of most of these achievements is the ability to process diamond at the micro- and nanoscale.
Novel fabrication methods enabled advanced diamond quantum devices such as micron-thin membranes for heterogeneous photonics~\cite{Riedel2017, Korber2023, Herrmann2023,  Yurgens2024,  Berghaus2025}, all diamond scanning probe sensors~\cite{Maletinsky2012, Appel2016, Zhou2017, Hedrich2020}, mechanical resonators~\cite{Tao2014, Khanaliloo2015a,  Momenzadeh2016, Heritier2018, Li2024, Joe2024}, and monolithic nanophotonic waveguides and photonic crystal (PhC) cavities~\cite{Evans2018, Machielse2019, Knaut2024, Ding2024}.
However, in several aspects, fabrication hurdles continue to limit the performance and scalability of these devices.
Specifically, obtaining high-quality, extended, and homogeneous arrays of devices with arbitrary geometries while maintaining high yield and integration potential remains a stumbling block.
Additionally, for color center-based devices, magnetic and charge fluctuators located on processed surfaces lead to a degradation of the spin and optical coherence of the color centers, emphasizing the need for well-terminated, atomically smooth diamond surfaces~\cite{Sangtawesin2019, Yu2025}.

The fabrication of released, free-standing devices is a key aspect of this challenge and several methods have emerged to address it~\cite{Toros2020, Mi2020, Rani2021}. Angled-etching~\cite{Burek2014, Latawiec2016, Atikian2017} and quasi-isotropic etching~\cite{Khanaliloo2015, Mouradian2017, Kuruma2024} rely on under-etching of structures fabricated on bulk diamond surfaces and work best for low-dimensional, thin structures or devices supported by a pedestal.
The downsides of either are the tendency to produce suboptimal surface quality (increased roughness) and the inability to obtain arbitrarily flat back surfaces, compromising the device performance~\cite{Ding2024}.
A different strategy to produce diamond membranes with thicknesses matching the device requirement proceeds via a ``smart-cut'' process combined with diamond epitaxy~\cite{Magyar2011, Aharonovich2012}. This method yields thin membranes, typically ranging from tens to a few hundred nanometers~\cite{Guo2021, Scheuner2024}, with higher surface quality compared to the underetching methods. However, the smart-cut approach is complex, resource-intensive, and the membranes are completely disconnected from the bulk diamond and therefore need to be rebonded to a carrier~\cite{Guo2024}.

\begin{figure*}[tbh]
\includegraphics[scale=1]{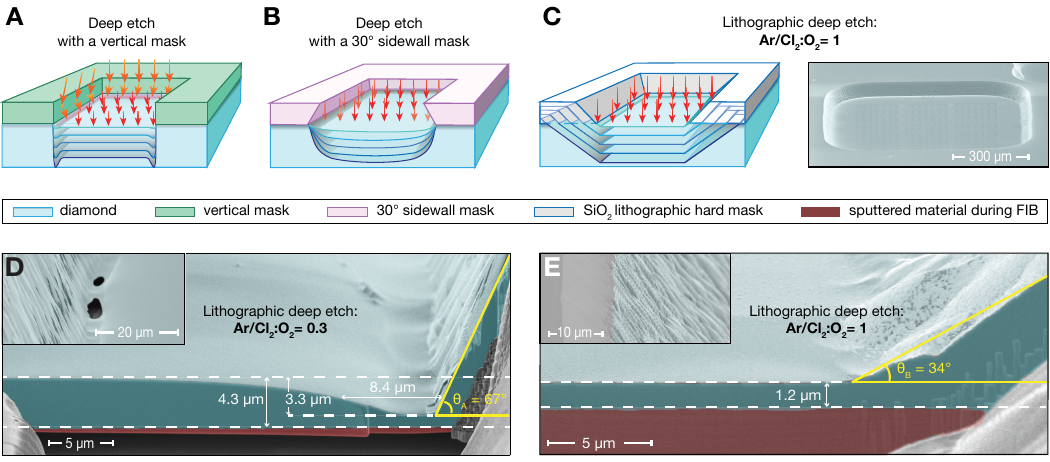}
\caption{
Diamond membrane processing.
(A) A vertical hard mask (green) and vertical diamond (blue) sidewall confine the etching plasma (orange arrows) at the edges of the etching pit, leading to the formation of trenches.
(B) A mask with $\sim$\ang{30} sidewalls (pink) decreases the etching plasma density (faded red arrows) at the edges of the etching pit, mitigating the formation of trenches but introducing thickness gradients in the membrane.
(C) Left: Evolution of the etching pit profile using an Ar/Cl\(_2\):O\(_2\) ratio of 1. The angled sidewall of the pit, created by etching the SiO\(_2\) mask during the Ar/Cl\(_2\) steps, mitigates plasma confinement and prevents trench formation. Right: SEM micrograph (taken at \ang{70} viewing angle) of a $\sim$\,45\,\textmu m deep etching pit. The resulting diamond membrane has an area of $\sim$\,760$\times$760\,\textmu m$^2$ and a thickness of 1.2\,\textmu m.
(D) SEM micrograph (same imaging angle) of the membrane thickness profile near the etching pit sidewall for an Ar/Cl\(_2\):O\(_2\) ratio of 0.3. The trench is 3.3\,\textmu m deep and 8.4\,\textmu m wide at half-depth. Inset: perforation of the membrane due to trenching.
(E) Same as (D), but for an Ar/Cl\(_2\):O\(_2\) ratio of 1. No trenches are visible. Inset: profile of the etching pit sidewall and of the SiO\(_2\) mask during etching. Terraces form due to different selectivity of the Ar/Cl\(_2\) and O\(_2\) plasma.}
\label{fig:fig1}
\end{figure*}

A more straightforward procedure for fabricating free-standing devices involves deep etching of a bulk diamond over an area defined by macroscopic shadow masks. By this approach, membranes can be created on millimeter scales, with low-roughness surfaces and a diamond quality limited only by the starting material~\cite{Appel2016, Heupel2022}. Yet, obtaining submicrometer thicknesses and low gradients across the entire etched membrane is highly challenging because of geometry-induced plasma enhancement effects~\cite{Hoekstra1998,Challier2018} that result in micrometers-deep trench build-ups along the membrane's perimeter (Fig.\,\ref{fig:fig1}\,A).
Reduction of the plasma flux near the sidewalls can be achieved using a shadow mask with $\sim$\ang{30} sidewalls (wider opening facing the diamond, Fig.\,\ref{fig:fig1}\,B), however, this method still produces significant thickness gradients that are dependent on the mask geometry~\cite{Momenzadeh2016,Challier2018}. Additionally, achieving accurate mask alignment and reliable mask fixation during the deep etch process remains challenging~\cite{Heupel2022}.

In addition to the ability to create thin structures, precise and efficient micro- and nanoscale patterning of diamond surfaces is essential for fabricating diamond-based quantum devices. Diamond’s hardness and chemical inertness demand mask materials with high chemical selectivity to transfer accurately the designed patterns~\cite{Toros2020}. Techniques such as electron beam lithography (EBL)~\cite{Burek2014, Mouradian2017} and focused ion beam (FIB) milling~\cite{Riedrich-Moller2012,Jung2019} have been successfully used to fabricate photonic crystal nanocavities and waveguides. However, while achieving high accuracy and resolution, these methods are resource-intensive and inefficient for large-scale patterning of micrometer-sized structures, such as scanning probe sensors~\cite{Appel2016} and platelets for heterogeneous photonic devices~\cite{Riedel2017}. Additionally, ion beam irradiation risks damaging the diamond surface and the embedded color centers. Optical lithography is a fast and scalable alternative, and it has already been used to pattern diamond structures~\cite{Zhou2017, Toros2018}. However, further development is needed to reproducibly reach the sub-micron resolution required to break out free-standing quantum devices by pick-and-place techniques.

In this work, we present first an improved fabrication strategy, the "lithographic deep etch" (LDE), to produce large-scale free-standing thin membranes. The LDE 1) eliminates problems with trench formation and thickness inhomogeneities, 2) enables high-accuracy positioning of the area to be deep-etched, and 3) allows us to define multiple etching windows of arbitrary shapes on one and the same diamond.
To demonstrate the potential of this method for quantum photonics, we pattern the front surface of bulk diamonds with micro- and nanostructures (diamond platelets for heterogeneous integration and PhC cavities, respectively) that are subsequently released from the back by the deep etch.
The free-standing structures we obtain have a sub-micron thickness (down to 70 nm) with a residual gradient limited only by the initial wedge of the diamond plate. They are compatible with pick-and-place transfer techniques while remaining supported by a frame robust enough for manual manipulation.
In a second part, we present a fast, simple, and scalable approach using optical lithography combined with an active feedback focusing system to define the front pattern of microstructures into diamond. To demonstrate its capability, we fabricate arrays of diamond platelets for heterogeneous integration and cantilevers for scanning‐probe sensors, achieving sub-micron resolution.

\begin{figure*}
\includegraphics[scale=1]{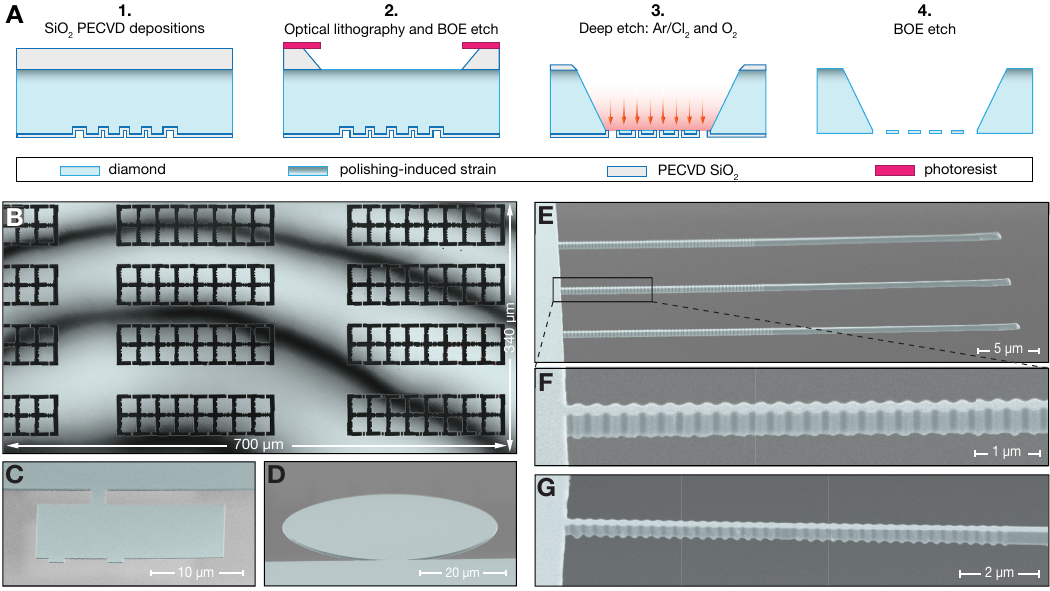}
\caption{
(A) Fabrication flow for the LDE process. (1) After the front pattern definition, the front surface is protected with a thin layer of SiO\(_2\) and the thick deep etch mask (10-22\,\textmu m of SiO\(_2\), light grey) is deposited on the backside. (2) Definition of the etching windows via optical lithography and a BOE etch to transfer the pattern to the hard mask. (3) Photoresist removal and ICP-RIE deep etch using Ar/Cl\(_2\) and O\(_2\) to release the nanostructures. (4) BOE etch to remove the hard mask.
(B) Laser scanning confocal microscope ($\lambda$=404 nm) image of a 700$\times$340\,\textmu m$^2$ region of the deep-etched side of a 2$\times$1\,mm$^2$ free-standing membrane patterned with platelets. The membrane has a wedge less than 0.6\,nm/\textmu m along the highest thickness gradient direction. The thickness difference between fringes is 84\,nm.
(C) SEM micrograph of a single 70\,nm thick free-standing platelet.
(D) SEM micrograph of a 50\,\textmu m diameter disk with a thickness of 500\,nm.
(E) SEM micrograph of 360\,nm thick free-standing tapered waveguides with distributed Bragg reflectors (DBR).
(F) Zoom in on the DBR and anchor point to the holding structure of one of the waveguides shown in (E).
(G) SEM micrograph of a 350\,nm thick free-standing tapered PhC cavity. All SEM images were obtaining at a 70$^{\circ}$ viewing angle.}
\label{fig:fig2}
\end{figure*}

The LDE process for creating large-scale free-standing membranes from a bulk diamond ($\sim$\,50\,\textmu m) relies on creating a thick SiO\(_2\) lithographic mask with outward-slanted sidewalls that retract as the deep etch progresses (Fig.\,\ref{fig:fig1}\,C). Our inductively coupled plasma reactive ion etching (ICP-RIE) recipe relies on alternating Ar/Cl\(_2\) and O\(_2\) plasmas~\cite{Maletinsky2012}. The former mostly leads to a smoothening of the diamond surface while the latter produces a fast, anisotropic etch of the diamond. Crucially, Ar/Cl\(_2\) also etches aggressively SiO\(_2\). Thus, tuning the etching time ratio of the plasmas allows for a controlled lateral retraction of the SiO\(_2\) mask. As a result, the angle of the etching pit's sidewalls can be tuned, preventing plasma confinement and the formation of trenches. The mechanism that avoids the trench formation during the LDE is shown in Fig.\,\ref{fig:fig1}\,C, together with a typical 45\,\textmu m deep etching pit resulting from our process, here with an area of 760$\times$760\,\textmu m$^2$ . The diamond membrane is 1.2\,\textmu m  thick.
To characterize how the membrane thickness profile close to the diamond sidewall evolves as a function of the Ar/Cl\(_2\):O\(_2\) content, we mill an inspection slot in the proximity of the etching pit sidewall via focused ion beam (FIB) to access the trench profiles. In Fig.\,\ref{fig:fig1}\,D, where a diamond pit etched with an Ar/Cl\(_2\):O\(_2\) ratio of 0.3 is shown, a trench with a depth of 3.3\,\textmu m and a width of 8.4\,\textmu m at half its depth is measured. The angle of the diamond etching pit sidewall from the sample plane is 67$^{\circ}$, showing that the plasma confinement, although mitigated, is still present and induces higher etching rates along the membrane perimeter and eventually leads to perforation (see inset of Fig.\,\ref{fig:fig1}\,D).
Increasing the Ar/Cl\(_2\):O\(_2\) ratio to one decreases the diamond etching pit sidewall angle close to the membrane to 34$^{\circ}$ (Fig.\,\ref{fig:fig1}\,E), while the sidewall has an average angle of $\sim$45$^{\circ}$ (inset Fig.\,\ref{fig:fig1}\,E). At this sidewall angle, the reflected ions at grazing angles no longer reach the diamond membrane but hit the sidewall such that no trench build-up is observed. In the inset of Fig.\,\ref{fig:fig1}\,E the diamond terraces, created by alternating the two plasmas, as well as the SiO\(_2\) lithographic mask are visible.

To integrate patterns needed for quantum devices into the diamond membrane, two approaches can be employed. In the first approach, the design ("front lithography") is patterned prior to membrane release, while in the second, patterning is performed after the creation of a thin diamond membrane. For the fabrication of sub-micron free-standing diamond nanostructures, we adopt the first approach to avoid potential damage to the fragile membrane during lithography.
The fabrication process for the LDE following the front pattern definition is illustrated in Fig.\,\ref{fig:fig2}\,A.
First, we use plasma-enhanced chemical vapor deposition (PECVD) to deposit a layer of SiO\(_2\) (50-100\,nm) to protect the front design, flip the sample, and deposit a thick (10-22\,\textmu m) layer of SiO\(_2\) as hard mask for the deep etch.
Second, optical lithography is performed to define the etching windows (see SI) and the design is transferred to the hard mask via a buffered oxide etchant (BOE) 10:1 etch. The isotropic wet etch guarantees outward-slanted sidewalls, with angles much lower than $\sim$\ang{45} near the diamond surface. This enables the hard mask to retract in the third step, in which the diamond is etched with alternating Ar/Cl\(_2\) and O\(_2\) plasmas in an ICP-RIE reactor. 
Fourth, once the desired membrane thickness is achieved, the mask is removed with a BOE (10:1) etch.

We characterize the thickness gradient of the free-standing membranes using a laser scanning confocal microscope (Keyence VK-X1100, $\lambda$=404\,nm). The measured wedge is consistent with that observed for the diamond plate before the LDE (see SI Fig.\,S3) and is solely determined by the intrinsic thickness gradient introduced during laser slicing and polishing. Thus, unlike deep etching approaches based on bulk masks~\cite{Challier2018, Heupel2020, Fuchs2021, Heupel2022}, the LDE process does not introduce any additional thickness inhomogeneities.
We quantify the wedge within a 700$\times$340\,\textmu m$^2$ region of a free-standing membrane patterned with platelets in Fig.\,\ref{fig:fig2}\,B knowing that the thickness difference between two destructive interference fringes is $\lambda / 2 n_d$=84\,nm, where $n_d$ is the refractive index of diamond. The entire membrane has a wedge below 0.6\,nm/\textmu m along the direction of the highest thickness gradient.
By characterizing the thickness gradient of the diamond plate before patterning devices, owing to the lithographic nature of the LDE, one can exploit the intrinsic wedge for tuning the thickness of the quantum devices.
Over a 1\,mm$^2$ region, we measured wedges down to 0.35\,nm/\textmu m along the highest thickness gradient direction (see SI Fig.\,S3), which projected on the typical size of our Fabry-Pérot microcavity platelets \cite{Riedel2017, Yurgens2024}, 20$\times$20\,\textmu m$^2$, leads to a maximal thickness difference of 7\,nm.

\begin{figure}
    \includegraphics[scale=1]{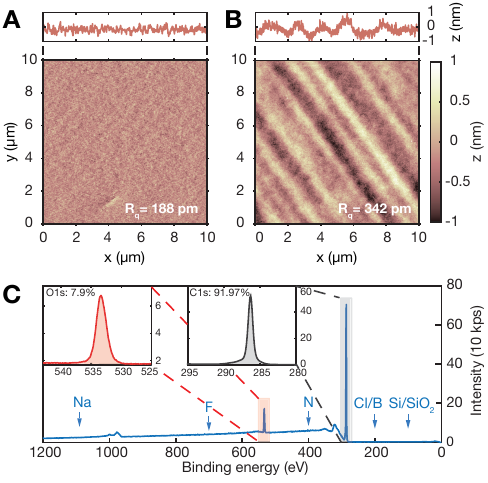}
    \caption{(A) AFM scan of the stress-relieved front diamond surface over an area of 10$\times$10\,\textmu m$^2$. The measured RMS roughness ($\mathrm{R_q}$) is 188\,pm. No increase of $\mathrm{R_q}$ is observed with the size of the AFM scan (see Supporting Information (SI) Fig.\,S1\,D).
    (B) AFM of the deep-etched diamond surface of a $\sim$900\,nm thin platelet over an area of 10$\times$10\,\textmu m$^2$. The $\mathrm{R_q}$ is 342\,pm, while for an area of 1$\times$1\,\textmu m$^2$ is 198\,pm. The higher surface roughness is due to different polishing performed on the two surfaces (see SI Fig\,S1\,E).
    (C) XPS survey measurement of the deep-etched surface of a diamond membrane after the LDE. Inset: high-resolution scans of the C1s and O1s peaks. The diamond has no contamination and is oxygen terminated after the triacid cleaning with approximately a monolayer of coverage.}
    \label{fig:fig3}
\end{figure}

The absence of a trench and the flexibility of the LDE process enable the fabrication of free-standing diamond structures with arbitrary thickness and size. As visible in Fig.\,\ref{fig:fig2}\,C, extended free-standing areas as thin as 70\,nm (see SI Fig.\,S4) can be realized, demonstrating this method as a highly competitive alternative to smart-cut techniques. Moreover, the LDE process enables the creation of flat structures connected to a supporting frame, and with arbitrary in-plane aspect ratios -- demonstrated in Fig.\,\ref{fig:fig2}\,D -- which are beyond the reach of conventional angled-etching or quasi-isotropic etching methods.
To demonstrate the potential of this approach, we fabricate some of the most demanding structures required for diamond-based quantum technologies, where stringent control over morphology and roughness of the back surface is critical. Specifically, we realize two distinct types of free-standing nanostructures, each with a thickness of 350–360\,nm: tapered waveguides with distributed Bragg reflectors (DBRs) and tapered photonic crystal (PhC) cavities. Both designs are optimized for a target wavelength of 1042\,nm, corresponding to the nitrogen-vacancy (NV) center's singlet transition, with an ideal thickness of 355\,nm. The full length of three tapered waveguides with DBRs is presented in Fig.\,\ref{fig:fig2}\,E, while a magnified view highlighting the DBR structure and the anchor points to the holding framework is shown in Fig.\,\ref{fig:fig2}\,F. Additionally, Fig.\,\ref{fig:fig2}\,G displays an SEM image of the two DBRs forming a PhC cavity, further illustrating the precision and viability of our fabrication process.

We characterize the surface morphology of the deep-etched side of the diamond nanostructures using atomic force microscopy (AFM). To facilitate this, free-standing diamond platelets, approximately 900\,nm thick, were released and bonded to a silicon chip via van der Waals forces, with the deep-etched surface facing upward (see SI Sec.IV). The morphology and roughness of this surface are compared to those of the front surface, which only underwent a 4\,\textmu m stress-relief etch to remove crystal damage introduced by the polishing~\cite{Appel2016}.
It is important to note that the plasma chemistry used for the deep etch closely mimics the one employed to achieve state-of-the-art stress-relieved diamond surfaces, utilizing alternating Ar/Cl\(_2\) and O\(_2\) plasmas (see SI Sec.I).
Fig.\,\ref{fig:fig3}\,A presents an AFM scan of the front surface after stress relief, revealing a remarkably low surface roughness (R$_q$) of 188\,pm, even over an area of 10$\times$10\,\textmu m$^2$. Notably, there is no significant increase in R$_q$ when comparing a 1$\times$1\,\textmu m$^2$ scan (187\,pm) to a 10$\times$10\,\textmu m$^2$ scan (see SI Fig.\,S1\,D).
This long-range flatness over tens of square micrometers is attributed to the combination of the stress-relief etching process and an improved diamond polishing technique developed by Almax easyLab (Belgium), which already yields surfaces with a roughness of approximately 200\,pm after polishing (see SI Fig.\,S1\,B). The exceptional surface quality and flatness of the released diamond micro- and nanostructures hold promise for photonics and quantum device applications, as polishing-induced surface waviness is a major limiting factor in achieving high Q-factors in diamond-based Fabry-Pérot microcavities~\cite{Flagan2022}.
Fig.\,\ref{fig:fig3}\,B shows a 10$\times$10\,\textmu m$^2$ AFM scan of the deep-etched surface, revealing a surface roughness of 342\,pm. While this initially suggests an increase in roughness due to the deep etch, it is crucial to consider that the back surface, before deep etching, had undergone a different polishing process performed by Almax easyLab, resulting in a significantly higher initial roughness of 976\,pm over a 10$\times$10\,\textmu m$^2$ area (see SI Fig.\,S1\,A). To allow a fair comparison, we applied our standard stress‐relief etch to a diamond sample that underwent the same initial polishing as the deep-etched surface.
The resulting R$_q$ value of 587\,pm indicates that the deep etch effectively smoothens the waviness induced by polishing, even more so than the stress-relief etch (see SI Fig.\,S1\,C). Finally, it is worth noting that polishing-induced waviness becomes relevant only for surface areas larger than 1$\times$1\,\textmu m$^2$, as at that scale, R$_q$ remains below 200\,pm for all considered cases.

The deep-etched back surface was further analyzed using X-ray photoelectron spectroscopy (XPS) with a beam size of 250\,\textmu m$^2$ and compared to the front surface.
The corresponding XPS survey shown in Fig.\,\ref{fig:fig3}\,C, is identical to that of a stress-relieved surface, exhibiting no detectable contamination. The high-resolution scan of the O1s peak confirms oxygen termination, with a coverage of approximately one monolayer, consistent with the expected outcome of the triacid cleaning process (refluxing mixture of concentrated perchloric, nitric, and sulfuric acids)~\cite{Sangtawesin2019}. Additionally, the high-resolution scan of the C1s peak shows no detectable low-energy shoulder, indicating a minimal presence of sp$^2$ carbon bonds. These bonds are known to lead to deep traps that in turn induce surface magnetic and charge noise, which can degrade the spin and optical coherence of shallow optically active spin defects in diamond~\cite{Sangtawesin2019}. The absence of contaminants and the low content of sp$^2$ carbon-related deep traps confirm the state-of-the-art surface quality achieved through the LDE process.

\begin{figure}
\includegraphics[scale=1]{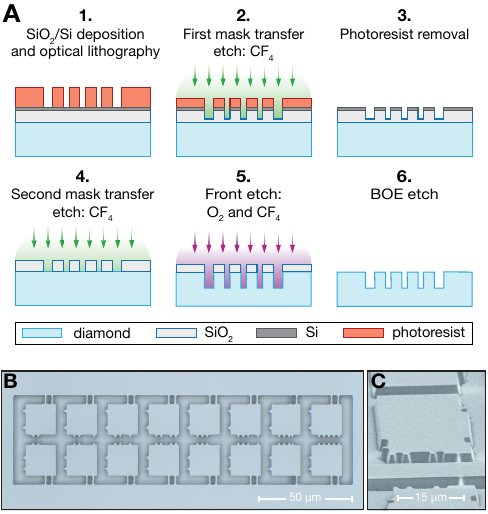}
\caption{(A) Fabrication flow for the DLW optical lithography and front etching of diamond micro-nanostructures. (1) PECVD deposition of the etch mask (300\,nm of SiO\(_2\)) and of 60\,nm Si layer for focusing the writing laser. After spin coating the photoresist, DLW optical lithography is performed. (2) Pattern transfer to the SiO\(_2\)/Si layers through a CF\(_4\) ICP-RIE etch. The etch is calibrated to stop a few tens of nm before etching through the SiO\(_2\) layer to avoid contamination and micro-masking of the diamond surface. (3) Photoresist removal through acetone and piranha cleaning. (4) The remaining few tens of nm of SiO\(_2\) and the Si layer are removed with CF\(_4\) (5) Pattern transfer to the diamond front surface via O\(_2\) and CF\(_4\) etches. (6) SiO\(_2\) removal with a BOE etch.
(B) Micrograph after pattern definition of 20$\times$20\,\textmu m$^2$ platelets with 1\,\textmu m wide bridges via optical lithography.
(C) SEM micrograph after pattern definition of 20$\times$40\,\textmu m$^2$ cantilevers with 200\,nm wide bridges via optical lithography.}
\label{fig:fig4}
\end{figure}

Next, we focus on the optical lithography process we developed to define efficiently the front pattern of the diamond microstructures. This method relies on focusing the laser of a direct laser writing (DLW) system onto the photoresist, thereby writing the desired pattern, and transferring the pattern to a SiO\(_2\) hard mask that ensures chemical resistance to O$_2$ plasma etching, thereby maintaining high pattern fidelity throughout the process. The fabrication process is illustrated in Fig.\,\ref{fig:fig4}\,A.
We begin with the PECVD deposition of a 300\,nm SiO\(_2\) etch mask and a 60\,nm Si layer to facilitate laser focusing.
Due to the transparency of both the diamond plate and the SiO\(_2\) layer, the Si layer is essential to provide the optical contrast needed to focus reliably the feedback laser in the DLW system on the top surface.
The DLW optical lithography is performed using a positive photoresist (see SI Sec.\,II\,A), and the pattern is transferred to the SiO\(_2\)/Si layers via a CF\(_4\) plasma ICP-RIE etch.
To prevent contamination and micromasking of the diamond surface, the etch is calibrated so that it stops a few tens of nm before reaching the diamond, allowing for photoresist removal. The remaining SiO\(_2\) and Si focusing layers are then fully removed by a final CF\(_4\) etch, completing the definition of the etch mask.
The pattern is subsequently transferred into the diamond using a O\(_2\) and CF\(_4\) plasma ICP-RIE etch, after which the mask is removed via a BOE etch. AFM and XPS measurements confirm that this process does not introduce any contamination or increase the surface roughness.

To evaluate our fabrication process, we pattern a 4$\times$4\,mm$^2$ area of the diamond surface with 20$\times$20\,\textmu m$^2$ platelets (see SI Fig.\,S2), each connected to the holding structure by 1\,\textmu m-wide bridges, compatible with pick-and-place transfer into a Fabry-Pérot optical microcavity~\cite{Yurgens2024}. A portion of the patterned area is shown in Fig.\,\ref{fig:fig4}\,B. Additionally, to assess the minimum feature size achievable using positive-tone lithography, we fabricate cantilevers for scanning NV magnetometry~\cite{Thiel2019}, achieving bridge widths as small as 200\,nm (Fig.\,\ref{fig:fig4}\,C).
By leveraging optical lithography with a positive-tone resist and a robust hard-mask transfer technique, we overcome the major limitations of conventional EBL-based approaches, enabling sub-micron resolution over millimeter-scale areas with significantly reduced processing time.

In conclusion, we have demonstrated the fabrication of large-scale, homogeneous, sub-micron thick, free-standing nano- and microstructures from single-crystal bulk diamond, tailored for integrated quantum technologies. A key challenge in deep-etching processes, trench build-up caused by plasma confinement near the mask and diamond etching pit sidewalls, was systematically investigated and mitigated through a lithographic SiO$_2$ mask. As a result, we achieved lithographically defined, millimeter-scale membranes with thickness gradients as low as 0.35\,nm/\,\textmu m, limited only by the initial wedge of the diamond plate after polishing. This enables the deterministic release of highly uniform nano- and microstructures with thicknesses down to 70\,nm. Comprehensive surface characterization confirms the exceptional quality of the fabricated structures, with atomically smooth, contamination-free surfaces ($\mathrm{R_q}$\,<\,200\,pm), ideal for hosting coherent spin defects. Additionally, the resulting nano- and microstructures are compatible with pick-and-place transfer techniques, facilitating integration with heterogeneous material platforms. 

To further advance diamond microfabrication, we developed a refined patterning approach based on DLW optical lithography, significantly reducing writing times, simplifying fabrication equipment, and enabling scalable, high-resolution structuring over large areas, making it a promising candidate for large-scale quantum device fabrication. These advancements mark a significant step toward scalable and high-performance diamond-based quantum technologies.

\begin{acknowledgments}
We thank Gerard Gadea Diez for fruitful discussions regarding the Si and the thick SiO$_2$ PECVD depositions, Marcus Wyss for the FIB milling, Kexin Wu for his contribution to the lithographic design, and the Swiss Nanoscience Institute for providing and maintaining much of the fabrication infrastructure. 
We acknowledge financial support by the ERC consolidator grant project QS2DM, and by the SERI Swiss quantum transitional call project ``QuantumLeap''.
A.C. acknowledges financial support from the Quantum Science and Technologies at the European Campus (QUSTEC) project of the European Union’s Horizon 2020 research and innovation program under the Marie Skłodowska-Curie grant agreement No.\,847471.
Y.Z., C.A.J.C. and C.G. acknowledge funding from the Swiss National Science Foundation, grant No.\,198898 and No.\,204036.
\end{acknowledgments}

\bibliography{referencesPaper}

\end{document}


\title{Supporting Information for:\\Homogeneous Free-Standing Nanostructures from Bulk Diamond\\ over Millimeter Scales for Quantum Technologies}

\date{June 11, 2025}

\maketitle

In this supporting information, we provide details for the developed fabrication processes, which can be broken down into three main steps:
\begin{enumerate}[label=\Roman*.]
    \item \hyperref[sec:sre]{Sample preparation and stress relief etch (SRE)} 
    \item \hyperref[sec:front]{Front pattern definition}
    \item \hyperref[sec:lde]{Lithographic deep etch (LDE)}
\end{enumerate}
Unless specified, all dry etch processes are done using inductively-coupled reactive ion etching (ICP-RIE) in a system from Sentech (SI 500), with the diamonds glued on silicon chips with Crystalbond 509. Etch parameters are listed in Table \ref{tab:table1}. Wet etching is performed either in a buffered oxide etchant (BOE) 10:1, in a piranha solution (sulfuric acid and hydrogen peroxide, 3:1) or in a refluxing mixture of concentrated perchloric, nitric, and sulfuric acids (triacid, 1:1:1). To ensure that the processes do not lead to contamination or morphological damage, surface quality is assessed at regular intervals using X-ray photoelectron spectroscopy (XPS) and atomic force microscopy (AFM), following Ref.\,\cite{Sangtawesin2019}.

\begin{table}[h!]
\caption{ICP-RIE plasma parameters for the SRE, front pattern definition, and the LDE. The mask transfer and front etch are carried out using a (6-inch) silicon carrier inside the reactor, while the SRE and LDE, which involves Ar/Cl$_2$ plasma, were carried out using a ceramics carrier wafer.}\label{tab:table1}
\begin{ruledtabular}
\begin{tabular}{cccccc}
 Plasma & ICP power & RF power/bias & Flux & Pressure & Etch rate \\
 & (W) & (W) & (sccm) & (Pa) & (nm/min) \\
\hline
Ar/Cl$_2$\footnotemark[1] & 400 & 100 & 25/40 & 1 & diamond: 40, SiO\(_2\): 90 \\
O$_2$\footnotemark[1] & 700 & 50 & 60 & 1.3 & diamond: 180, SiO$_2$: 2 \\
CF$_4$\footnotemark[2] & 50 & 45 & 30 & 0.4 & Si: 50, SiO$_2$: 40, photoresist: 60\\
O$_2$\footnotemark[3] & 500 & 100 & 50 & 0.5 & diamond: 160 \\
O$_2$\footnotemark[4] & 400 & 200 & 30 & 1.3 & diamond: 60 \\
\end{tabular}
\end{ruledtabular}
\footnotetext[1]{SRE and LDE}
\footnotetext[2]{Mask transfer etch and front etch-- optical lithography}
\footnotetext[3]{Front etch -- optical lithography}
\footnotetext[4]{Front etch -- e-beam lithography}
\end{table}

\section{Sample preparation and stress relief etch}\label{sec:sre}
\begin{figure}
\includegraphics[scale=1]{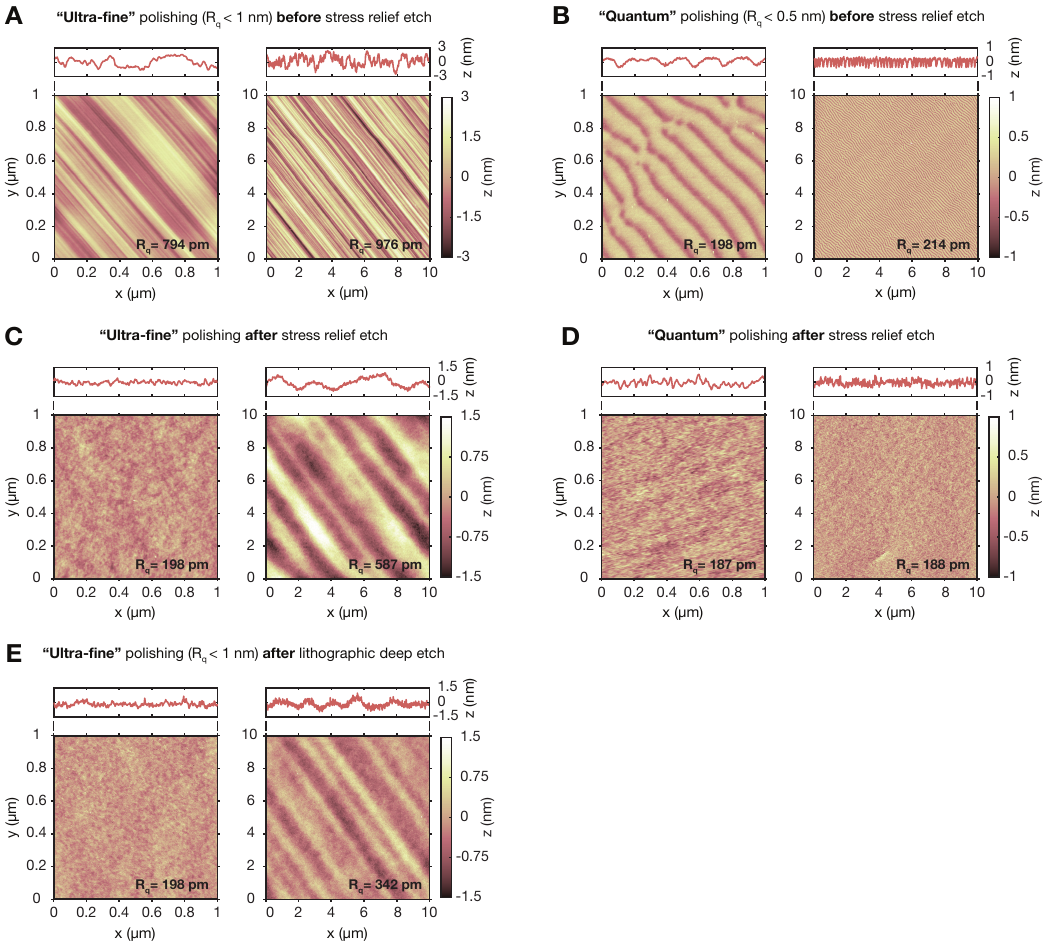}
\caption{\qtyproduct[product-units = power]{1 x 1}{\um} and \qtyproduct[product-units = power]{10 x 10}{\um} AFM scans of the diamond surface for the two different polishing methods "Ultra-Fine Polishing" and "Quantum Polishing" performed by Almax easyLab at different processing steps.}
\label{fig:fig5}
\end{figure}

We fabricate our samples from commercially available synthetic single-crystal diamond plates grown by chemical vapor deposition (CVD). Sample A is a \qty{50}{\um}-thick optical grade (100) diamond plate, double-side polished to R$_q <$\qty{3}{\nm} ([N$_S$] <\,1 ppm, [B] <\,5 ppb, Delaware Diamond Knives). Samples B, C and D are electronic grade (100) plates ([N$_S$] <\,5 ppb, [B] <\,1 ppb, Element Six), that were laser-diced into \qty{50}{\um}-thick membranes (Almax easyLab).
Samples B and C are polished on the front side to R$_q <$\qty{0.5}{\nm} ("Quantum polishing", Almax easyLab) and on the back to R$_q <$\qty{1}{\nm} ("Ultra-fine polishing", Almax easyLab). Sample D is double-sided polished to R$_q <$\qty{1}{\nm} ("Ultra-fine polishing", Almax easyLab). Before any process, the samples were cleaned in triacid. The AFM characterization reveals that the “Ultra-fine polishing” method achieves a peak-to-peak waviness of \qty{6}{nm} and a typical surface roughness of R\textsubscript{q}\,=\,\qty{794}{\pm} over an area of \qtyproduct[product-units = power]{1 x 1}{\um}, which further increases to R\textsubscript{q}\,=\,\qty{976}{\pm} over an area of \qtyproduct[product-units = power]{10 x 10}{\um} (Fig.\,\ref{fig:fig5}\,A). In contrast, the “Quantum polishing” method exhibits a dramatically reduced peak-to-peak waviness of \qty{1.5}{nm}, and the surface roughness remains nearly constant when the analyzed area increases from \qtyproduct[product-units = power]{1 x 1}{\um} (R\textsubscript{q}\,=\,\qty{198}{\pm}) to \qtyproduct[product-units = power]{10 x 10}{\um}, with a value as low as R\textsubscript{q}\,=\,\qty{214}{\pm} (Fig.\,\ref{fig:fig5}\,B). The low waviness and exceptional smoothness achieved through “Quantum polishing” hold significant promise for fabricating diamond surfaces that can increase the performances of diamond photonic devices \cite{Flagan2022}. The \qty{50}{\um}-thick diamond plates used to realize the free-standing structures exhibit a wedge between \qty{0.25}{\um} and \qty{2.5}{\um} across the \qty{4.5}{\mm} lateral dimension after laser slicing and polishing. As an example, sample C exhibits wedges across the 4 sides ranging between 0.25 and \qty{1.4}{\um}.

Sub-surface crystal damage and strain induced by mechanical polishing can extend micrometers-deep~\cite{Volpe2009, Friel2009, Section2012}, and is extremely detrimental to the color centers embedded in the material~\cite{Yurgens2022, Sangtawesin2019, Rondin2014}. A damage-free lattice is recovered by etching away $\sim$\qty{4}{\um} of diamond from the front surface using Ar/Cl\(_2\) and O\(_2\) plasma.
Starting this stress relief etch (SRE) with an extended Ar/Cl\(_2\) plasma step is crucial: chlorine chemistry removes the damaged layer without propagating lattice defects into the crystal \cite{Friel2009}. A last O\(_2\) step removes possible deleterious Cl\(_2\) contamination~\cite{Tao2014}. The recipe used for all samples was 8 cycles of (300 s Ar/Cl$_2$ + 150 s O$_2$) with the plasma parameters specified in Table \ref{tab:table1}.
The AFM characterization of the "ultra-fine" polished surface shows significantly reduced surface waviness and roughness after the SRE, with the peak-to-peak waviness decreasing by a factor of four to \qty{1.5}{\nm} and a roughness (R\textsubscript{q}) of \qty{198}{\pm} on a \qtyproduct[product-units = power]{1 x 1}{\um} area, and by nearly a factor of two to \qty{3}{\nm} and R\textsubscript{q} of \qty{587}{\pm} on a \qtyproduct[product-units = power]{10 x 10}{\um} area (Fig.\,\ref{fig:fig5}\,C). The smoothing effect of the etching process is less pronounced on the "quantum" polished surface, where the surface roughness slightly decreases to R\textsubscript{q} = \qty{188}{\pm} and remains constant up to an area of \qtyproduct[product-units = power]{10 x 10}{\um}, while the waviness becomes more randomized (Fig.\,\ref{fig:fig5}\,D).
Although the differences in surface waviness and roughness between the two polishing processes appear negligible over a \qty{1}{\um\squared} area, the improvements in morphology and roughness are significant over larger areas spanning several \unit{\um\squared}.

An XPS characterization is performed to check for the most common diamond contaminants (B, F, Na, Si/SiO$_2$, Cl). An XPS survey spectrum after the stress relief etch is shown in Fig.\,3\,C of the main text, showing no detectable amounts of contaminants. The detected oxygen is the one terminating the diamond surface after the triacid cleaning.

\section{Front pattern definition}\label{sec:front}
In this work, we fabricate two main types of free-standing structures: diamond platelets -- used either to functionalize Fabry-Pérot microcavities or as cantilevers for scanning NV magnetometry (samples B and D) -- and photonic crystal (PhC) cavities and waveguides with a distributed Bragg reflector (DBR) designed for the singlet transition at \qty{1042}{\nm} of the NV center (sample C). The first type consists of 20-\qtyproduct[product-units = power]{40 x 20}{\um} diamond platelets, attached to a holding bar by \qty{5}{\um} long and 0.2-\qty{1}{\um} wide bridges. The second type, PhC cavities and waveguides, have a width of 535–635 nm, a DBR periodicity of 265–300 nm, and a tapered section ending at a thickness of 50 nm. For the fabrication of PhC cavities and waveguides, we employ a well-established process based on electron beam lithography (EBL) using FOx-16 (DuPont, formerly Dow Corning) as a resist. In contrast, for the fabrication of diamond platelets, we develop a refined process based on optical lithography.

\subsection{Optical lithography: sample B and D}\label{sec:front,subsec:optical}

\begin{figure}[b]
\includegraphics[scale=1]{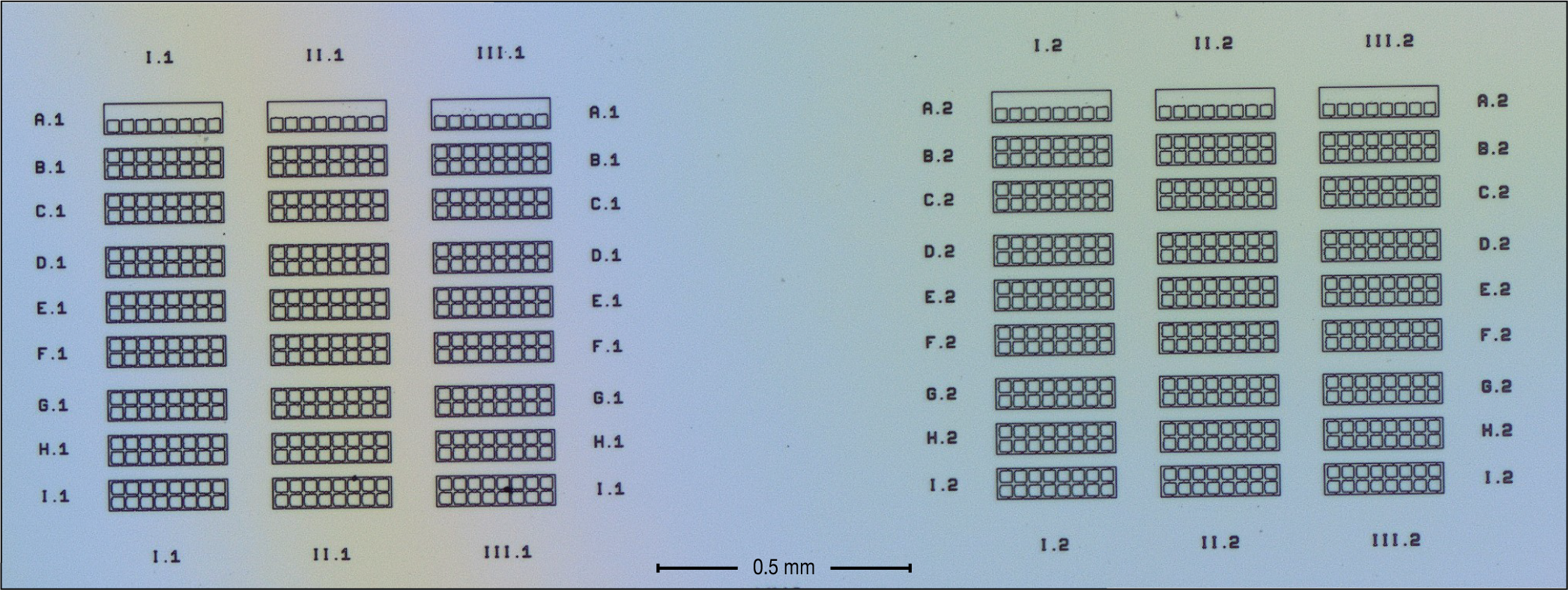}
\caption{
Micrograph of sample B after front pattern definition of two 1 mm$^2$ arrays of \qtyproduct[product-units = power]{20 x 20}{\um} platelets with \qty{1}{\um} wide bridges by optical lithography.}
\label{fig:fig6}
\end{figure}

The hard mask used for the front pattern definition via optical lithography consists of a 300 nm SiO$_2$ layer deposited through plasma-enhanced chemical vapor deposition (PECVD) at 300 $^{\circ}$C (Oxford Instruments, PlasmaPro 80 PECVD) on the diamond's front surface.
Optical lithography is performed with a direct laser writer (DLW) equipped with an all-optical auto-focus module (Heidelberg Instruments, uPG 101). Since the diamond plate and the SiO$_2$ layer do not have enough optical contrast to guarantee reliable focusing of the all-optical auto-focus module on the top surface, a 60 nm Si layer was deposited via PECVD at 300 $^{\circ}$C on top of the SiO$_2$ hard mask without breaking vacuum.
AR 300-80 New (Allresist GmbH) is applied as an adhesion promoter and S1813 (Shipley, Microposit) was chosen as photoresist, forming a 1.3 $\mu$m thick layer. The pattern is written using the all-optical auto-focus function of the DLW at 7.5 mW power, and developed with MF-319 (Shipley, Microposit).
The pattern is then transferred from the photoresist to the Si/SiO$_2$ mask using  ICP-RIE with a CF$_4$ plasma. The etching parameters are listed in Table \ref{tab:table1}. This process removes the Si layer and approximately 260 nm of the SiO$_2$ mask, allowing photoresist removal (with acetone and a piranha clean) without contaminating the diamond surface.
To complete the mask transfer etch, an additional CF$_4$ etch is performed to remove the remaining Si layer and the last $\sim$\qty{40}{\nm} of the hard mask. The pattern is then transferred from the SiO$_2$ hard mask to the diamond using an O$_2$ and CF$_4$ ICP-RIE process that etches approximately 3 $\mu$m of diamond material and effectively transfers the desired microstructure. The recipe used is 8 cycles of (10s CF$_4$ + 120s O$_2$). The detailed etching parameters for this process are provided in Table \ref{tab:table1}. The hard mask is finally removed with a BOE etch and the sample is cleaned using a triacid clean. In Fig.\,\ref{fig:fig6} a micrograph of two 1 mm$^2$ arrays of diamond platelets patterned on sample B is shown. AFM and XPS characterizations were performed to verify that none of the fabrication steps damaged the diamond surface.

\subsection{Electron beam lithography: sample C}\label{{sec:front,subsec:ebl}}
For front pattern definition via electron beam lithography, we use FOx-16 both as a resist and a hard mask, a common approach for fabricating diamond micro- and nanostructures \cite{Appel2016}. To improve adhesion between the resist and the diamond surface, and to prevent charge buildup during writing, a 20 nm Ti layer was deposited by electron beam evaporation (Alliance-Concept, EVA 760) onto the front surface of the diamond before spin coating. FOx-16 was then spin-coated, resulting in a mask thickness of approximately \qty{1.25}{\um}. The pattern was written using a 100 keV EBL system (Raith, EBPG5000), and the resist was developed in tetramethylammonium hydroxide (TMAH). The hard mask design was then transferred to the diamond surface using an O$_2$ ICP-RIE etch (Oxford Instruments, PlasmaPro100 Cobra), which etched about \qty{1.15}{\um} of diamond material. Finally, the FOx-16 mask was removed with BOE, and the sample was cleaned using a triacid cleaning process.

\section{Lithographic deep etch (LDE)}\label{sec:lde}
The lithographic deep etch process begins by protecting the front surface of the diamond with a 50-\qty{100}{\nm} SiO$_2$ layer, deposited via PECVD at 300 $^{\circ}$C. The sample is then flipped and a 10-\qty{22}{\um} SiO$_2$ layer is deposited on the backside.
Next, we define the etching windows of the hard mask by optical lithography using the same DLW system as for the front lithography. AR 300-80 New is applied as adhesion promoter and AZ 4562 (MicroChemicals GmbH) as photoresist, resulting in an \qty{8.8}{\um} thick layer that offers high etch resistance to BOE. For patterning the deep etch windows, we align the design to the front pattern that is visible through the mask and the diamond plate. As submicrometer resolution is not required, we manually set the writing distance and omit the Si focusing layer, which is used for the all-optical autofocus system. We use a laser power of 120 mW and develop the resist in AZ 400K 1:4 (MicroChemicals GmbH). The lithographic design is then transferred to the SiO$_2$ hard mask via wet etching in BOE with an etching rate of 200 nm/min. We then remove the photoresist using acetone and a piranha clean.

\begin{figure}[h!]
\includegraphics[scale=1]{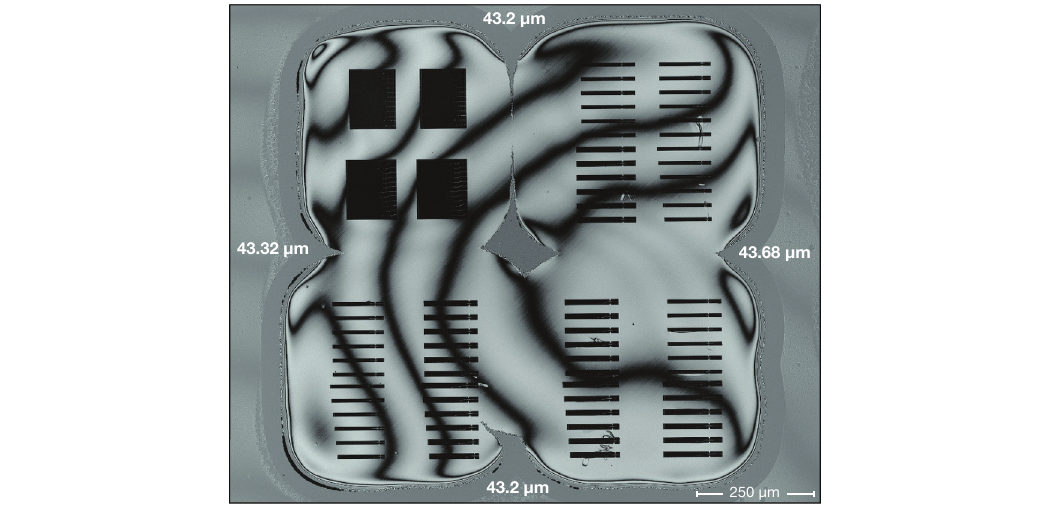}
\caption{
Laser scanning confocal microscope image (Keyence VK-X1100) showing a \qtyproduct[product-units = power]{1 x 1}{\mm} section of a free-standing membrane patterned with PhC and waveguides. The membrane exhibits a wedge of \qty{0.35}{\nm/\um} in the direction of the highest thickness gradient, with a thickness difference of \qty{84}{\nm} between fringes set by the laser wavelength \(\lambda = 404 \, \mathrm{nm}\). This value matches the wedge of the diamond plate measured on the 4 sides of the writefield via interferometric measurements with the same confocal microscope.
}
\label{fig:fig7}
\end{figure}

The isotropic wet etch of the lithographic hard mask produces a non-uniform sidewall angle, considerably less than 45$^{\circ}$ near the diamond surface. Combined with the selection of plasma gases for deep etching -- which gradually erodes the mask, causing it to retract concurrently with the diamond etch -- this phenomenon forms the basis of the process’s key principle: reducing plasma confinement and, consequently, the ion flux near both the mask and the diamond etch pit sidewalls. It is important to note that over-etching the deep etch mask creates steeper sidewall angles adjacent to the diamond surface, leading to the formation of a shallow trench that may eventually perforate the diamond membrane (see Fig.\,\ref{fig:fig7}).

Although CF$_4$ and SF$_6$ are effective in etching SiO$_2$, we opted for Ar/Cl$_2$, motivated by the excellent surface quality obtained after the SRE. Ar/Cl$_2$ also exhibits greater chemical selectivity for SiO$_2$ over diamond. O$_2$ plasma, on the other hand, shows a higher etch rate for diamond compared to SiO$_2$. By employing the same plasma chemistry and parameters as in the SRE process (see Table \ref{tab:table1}) -- Ar/Cl$_2$ and O$_2$ gases -- we create diamond terraces via preferential etching of either diamond or SiO$_2$ by switching between the two plasmas. This allows us to control the angle of the diamond etching pit sidewall by adjusting the Ar/Cl$_2$:O$_2$ ratio. To investigate the process, we used two different etching recipes:
\begin{itemize}
    \item sample A: 150 s Ar/Cl$_2$ + (90 s Ar/Cl$_2$ + 300 s O$_2$)$\times$N \quad  \quad  \quad --  \quad Ar/Cl$_2$:O$_2$ ratio $=0.3$
    \item samples B, C: 150 s Ar/Cl$_2$ + (150 s Ar/Cl$_2$ + 150 s O$_2$)$\times$N \quad --  \quad Ar/Cl$_2$:O$_2$ ratio $=1$
\end{itemize}
where N is the number of cycles needed to release the structures with the desired thickness. Similarly to the SRE, the deep etch always ends with an O$_2$ plasma step. Following the etch, the lithographic mask and front-side protection were removed through a BOE etch, and any residual contamination was eliminated by a triacid clean.
AFM and XPS characterizations were carried out to confirm that none of the fabrication steps damaged or contaminated the diamond surface, and the results are shown in Fig.\,\ref{fig:fig5}\,E and in Fig.\,3\,C of the main text. As visible from the peak-to-peak waviness and R$_q$ values of the \qtyproduct[product-units = power]{10 x 10}{\um} scans shown in Fig.\,\ref{fig:fig5}\,C and \ref{fig:fig5}\,E, the LDE smoothens even more the polishing-induced waviness than the SRE thanks to the longer exposure to Ar/Cl$_2$ plasma.

The thickness gradient of the diamond membranes after LDE was also characterized. To understand whether our deep etch strategy introduces further thickness inhomegenities in the fabricated free-standing membranes, we measured the thickness of the diamond plate via interferometric measurements using a confocal microscope (Keyence VK-X1100, \(\lambda = 404 \, \mathrm{nm}\)) around the writefields patterned with the micro- and nanostructures before the LDE process. We then compared the thickness difference across the 4 sides of the writefields, with the one that can be extracted from a laser scanning confocal image performed with the same microscope, as shown in Fig.\,\ref{fig:fig7}. Knowing that the thickness difference between fringes is \qty{84}{\nm} and counting the number of fringes, one can compare the thickness gradient of the patterned area before and after the deep etch. We show that the LDE process does not introduce further thickness inhomogeneities to the starting wedge of the diamond plate caused by laser slicing and polishing and the membrane exhibits a wedge of \qty{0.35}{\nm/\um} in the direction of the highest thickness gradient.

\begin{figure}[h!]
\includegraphics[scale=1]{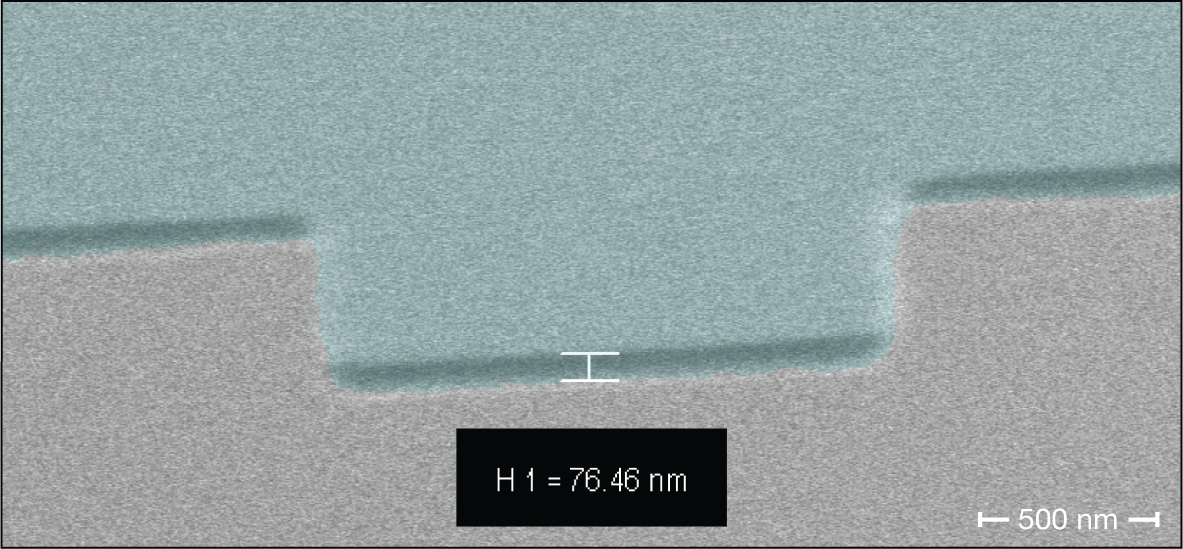}
\caption{
SEM micrograph at 70$^{\circ}$ viewing angle of a 70 nm thick platelet coated with 10 nm of Ti as a discharging layer. 
}
\label{fig:fig8}
\end{figure}

\newpage
Fig.\,\ref{fig:fig8} shows a 70 nm thick platelet fabricated with the LDE. It is coated with 10 nm of Ti for efficient discharging during the SEM investigation and measured at a viewing angle of 70$^{\circ}$. The height measurement in Fig.\,\ref{fig:fig8} does not take into account the viewing angle, and so the real height is 6\% larger than the measured one.

It has to be noted that while our deep etch fabrication flow can be applied to a large range of starting thicknesses of the diamond plate, for free-standing structures with sub-micrometer thickness we found that \qty{50}{\um} is a good compromise between mechanical stability, processing times, and thickness of the SiO$_2$ hard mask needed to release the structures with no trench along the etching pit sidewall.

Lastly, we have found that the LDE is also a powerful tool when it comes to releasing several micrometer-thick structures such as cantilevers for scanning NV magnetometry, as it allows us to deterministic release all devices in large arrays. Depending on the thickness of the structures to be released and the initial wedge of the diamond, trench formation is less of an issue, meaning that the LDE can be sped up using a lower Ar/Cl\(_2\):O\(_2\) ratio such as 0.3, which also significantly reduces the required thickness of the SiO$_2$ hard mask.

\section{Platelet release for AFM characterization}\label{sec:platelet_dep}

\begin{figure}[h!]
\includegraphics[scale=1]{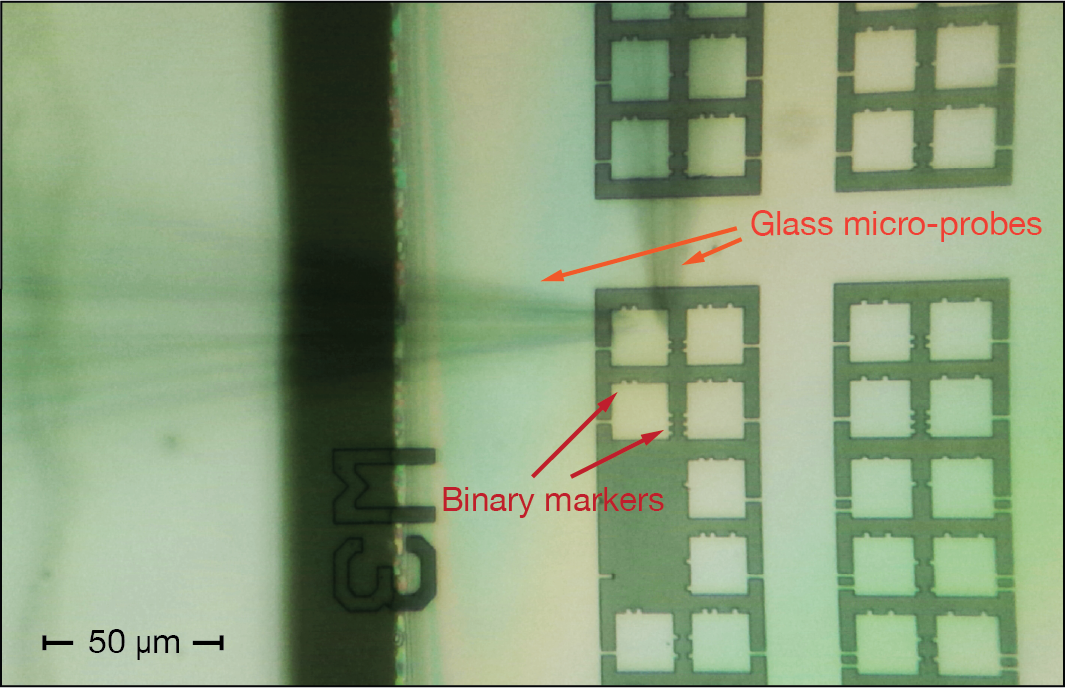}
\caption{
Platelet deposition with a micromanipulation station using glass micro-probes. The specific platelet is identified via binary markesrs that helps identifying also the front from the back surface.
}
\label{fig:fig9}
\end{figure}

To characterize the LDE back surface with AFM, we release some free-standing diamond platelets onto a Si chip using a home-assembled micromanipulation station. Glass microprobes, crafted with a glass needle puller, are utilized to snap the bridge connecting the platelet to the holding bar by applying pressure to the platelet's surface, as shown in Fig.\,\ref{fig:fig9}. The surface orientation of the platelet that lands on the silicon chip is determined using non-mirror-symmetric binary code markers added to the platelet (see Fig.\,\ref{fig:fig9}).
Once the back surface is correctly oriented facing upwards, we gently maneuver the platelet with the microprobes until van der Waals forces bond it to the Si chip, and it therefore cannot be moved anymore.

\bibliography{referencesPaper}